\documentclass[preprint, 12pt]{elsarticle}

\usepackage{amssymb}
\usepackage{lineno}
\usepackage{xcolor}
\usepackage{hyperref}
\definecolor{blue}{rgb}{0, 0.498, 0.675}
\hypersetup{breaklinks, colorlinks=true, citecolor=blue, linkcolor=blue, urlcolor=blue}

\journal{}

\usepackage[paperwidth=8.5in, paperheight=11in, margin=1in, headheight=0pt, headsep=0pt]{geometry}
\usepackage{enumitem}
\setitemize{leftmargin=12pt, itemsep=0pt}

\renewcommand\theaffn{}
\makeatletter
\def\printFirstPageNotes{%
  \iflongmktitle
    \let\columnwidth=\textwidth
  \fi
\ifdoubleblind
\else
  \ifx\@tnotes\@empty\else\@tnotes\fi
  \ifx\@nonumnotes\@empty\else\@nonumnotes\fi
  \ifx\@cornotes\@empty\else\@cornotes\fi
  \ifx\@elseads\@empty\relax\else
   \let\thefootnote\relax
   \footnotetext{\ifnum\theead=1\relax
      Corresponding author:\space\else
      \textit{Email addresses:\space}\fi
     \@elseads}\fi
  \ifx\@elsuads\@empty\relax\else
   \let\thefootnote\relax
   \footnotetext{\textit{URL:\space}%
     \@elsuads}\fi
\fi
  \ifx\@fnotes\@empty\else\@fnotes\fi
  \iflongmktitle\if@twocolumn
   \let\columnwidth=\Columnwidth\fi\fi
}

\gdef\emailauthor#1#2{\stepcounter{ead}%
     \g@addto@macro\@elseads{\raggedright%
      \let\corref\def Karr, Jonathan R. (\href{mailto:#1}{#1})}%
}

\long\def\pprintMaketitle{\clearpage
  \iflongmktitle\if@twocolumn\let\columnwidth=\textwidth\fi\fi
  \resetTitleCounters
  \def\baselinestretch{1}%
  \printFirstPageNotes
  \begin{\elsarticletitlealign}%
 \thispagestyle{pprintTitle}%
   \def\baselinestretch{1}%
    \Large\@title\par\vskip18pt%
    \ifx\@elsarticlenewpageafter\newpage@after@title
      \newpage
    \fi%
    \ifdoubleblind
      \vspace*{2pc}
    \else
      \normalsize\elsauthors\par\vskip10pt
      \footnotesize\elsaddress\par\vskip36pt
    \fi
    \ifx\@elsarticlenewpageafter\newpage@after@author
      \newpage
    \fi%
    \hrule\vskip12pt
    \ifvoid\absbox\else\unvbox\absbox\par\vskip10pt\fi
    \ifvoid\keybox\else\unvbox\keybox\par\vskip10pt\fi
    \hrule\vskip12pt
    \ifx\@elsarticlenewpageafter\newpage@after@abstract
      \newpage
    \fi%
    \end{\elsarticletitlealign}%
    \gdef\thefootnote{\arabic{footnote}}%
  }

\makeatother

\let\oldcite=\cite
\renewcommand{\cite}[1]{\textcolor{blue}{\oldcite{#1}}}

\let\oldcitep=\citep
\renewcommand{\citep}[2][]{\textcolor{blue}{\oldcitep[#1]{#2}}}

\usepackage[labelfont=bf]{caption}

\makeatletter
\renewcommand\@biblabel[1]{#1.}
\makeatother

\setcitestyle{numbers}
\setcitestyle{sort&compress}

\begin{document}


\begin{frontmatter}



\title{\textbf{Centralizing data to unlock whole-cell models}}

\author[sinai]{Yin Hoon Chew and Jonathan R. Karr}
\ead{karr@mssm.edu}

\affiliation[sinai]{organization={Department of Genetics and Genomic Sciences, Icahn School of Medicine at Mount Sinai},
             addressline={1425 Madison Avenue},
             city={New York},
             postcode={10029},
             state={NY},
             country={USA}}

\begin{abstract}
Despite substantial potential to transform bioscience, medicine, and bioengineering, whole-cell models remain elusive. One of the biggest challenges to whole-cell models is assembling the large and diverse array of data needed to model an entire cell. Thanks to rapid advances in experimentation, much of the necessary data is becoming available. Furthermore, investigators are increasingly sharing their data due to increased emphasis on reproducibility. However, the scattered organization of this data continues to hamper modeling. Toward more predictive models, we highlight the challenges to assembling the data needed for whole-cell modeling and outline how we can overcome these challenges by working together to build a central data warehouse.
\end{abstract}








\end{frontmatter}

\thispagestyle{empty}

\section*{Introduction}
More comprehensive and more predictive models of cells are broadly perceived as vital for understanding, controlling, and designing biology. For example, whole-cell models would likely help scientists conduct experiments in silico with unprecedented control and resolution \cite{carrera2015build}, help physicians precisely treat each patient's unique genomics \cite{tomita2001whole}, and help bioengineers rationally design synthetic cells \cite{marucci2020computer}.

Recently, scientists have taken several steps toward whole-cell models, producing large-scale models of \textit{Mycoplasma genitalium} \cite{karr2012whole, burke2020biochemical}, \textit{Mycoplasma mycoides} \cite{thornburg2019kinetic}, \textit{Escherichia coli} \cite{thiele2009genome, roberts2011noise, carrera2014integrative, macklin2020simultaneous}, \textit{Saccharomyces cerevisiae} \cite{munzner2019comprehensive, ye2020comprehensive}, and human epithelial cells \cite{ghaemi2020silico} among others. Researchers have also begun to explore how whole-cell models could help guide personalized medical decisions \cite{bordbar2015personalized} and design synthetic cells \cite{purcell2013towards, Rees2020}.

Despite substantial interest, whole-cell models remain elusive due to numerous challenges, including integrating vast information about diverse biochemical processes \cite{Takahashi2002}, accounting for the structure and organization of cells and their numerous components \cite{im2016, luthey2021integrating}; simulating \cite{goldberg2016toward}, calibrating \cite{babtie2017deal, stumpf2021statistical}, visualizing \cite{macklin2014future, feig2019whole}, and validating \cite{macklin2014future, feig2019whole} high-dimensional, computationally-expensive, hybrid models; and developing models collaboratively \cite{singla2021community, waltemath2016toward}. Toward a framework for whole-cell modeling, we and others have summarized these challenges \cite{macklin2014future, goldberg2018emerging, szigeti2018blueprint, feig2019whole}.

To help focus efforts to accelerate whole-cell modeling, we recently surveyed the community about the bottlenecks to progress \cite{szigeti2018blueprint}. Most respondents expressed that the main immediate barrier to more predictive models is insufficient experimental data and knowledge.

Undeniably, we do not yet have enough data to completely model a cell. As a result, complete models of entire cells are not presently feasible. Nevertheless, we believe that significantly more comprehensive models can already be constructed by leveraging the substantial data that is already available. Thus, in our opinion, the practical bottleneck to better models is not our limited experimental capabilities, but the scattered organization of our existing data. Furthermore, as our experimental capabilities continue to expand rapidly, we believe that it is critical to begin to develop whole-cell modeling capabilities now so that we are prepared to realize whole-cell models when sufficient data is available.

To focus efforts to address this bottleneck, here we explore the data that is already available and how we can best leverage it for whole-cell modeling. First, we outline the data that is needed for whole-cell modeling. Second, we highlight exemplary resources that already provide key data. Third, we assess the challenges to moving beyond these resources. Finally, we present a roadmap to assembling a data warehouse for whole-cell modeling. We firmly believe that such a warehouse would accelerate the development of more predictive models.

\section*{The mountain of data needed to model an entire cell}
Modeling an entire cell will likely require similarly comprehensive experimental data. At a minimum, this will likely include (a) the sequence of the cell's genome; (b) data about the structure of its genome, such as the location of each replication origin, promoter, and terminator; (c) information about the structure, abundance, turnover, and spatial distribution of each molecule in the cell; (d) information about each molecular interaction that can occur in the cell, including the molecules that participate in each interaction and the catalysis, rate, thermodynamics, and duration of each interaction; and (e) global information about the temporal dynamics and spatial organization of the cell, such as the organization of its life cycle, its size, shape, and subcellular organization.

\section*{The sea of data that could be repurposed for whole-cell modeling}
Compared to the experimental capabilities of an individual lab or even a consortium, this laundry list of data seems insurmountable. Without a quantum leap forward in automation or a massive increase in funding, we expect the data needed for whole-cell modeling to exceed the experimental capabilities of most labs for the foreseeable future. 

Although little data has been explicitly collected for whole-cell modeling, the scientific literature already contains substantial relevant data. Furthermore, much of this data is already publicly accessible due to an increasing culture of data sharing. Taken together, we believe that substantial data can be repurposed for more comprehensive models.

Exemplary data resources that we believe can be repurposed for whole-cell modeling include ECMDB \cite{sajed2016ecmdb}, YMDB \cite{pmid27899612}, PaxDB \cite{wang2015version}, PSORTdb \cite{pmid33313828}, BRENDA \cite{chang2021brenda}, and SABIO-RK \cite{wittig2018sabio}. ECMDB and YMBD contain thousands of measurements of the concentrations of metabolites in \textit{E. coli} and \textit{S. cerevisiae}. PaxDB contains over 1 million measurements of the abundances of proteins in over 50 organisms. PSORTdb contains over 10,000 measurements of the localization of proteins in over 400 organisms, as well as predicted localizations for over 15,000 organisms. Together, BRENDA and SABIO-RK contain over 300,000 kinetic parameters for thousands of metabolic reactions. In our experience, BioNumbers \cite{milo2010bionumbers} is also a valuable collection of data for modeling.

In addition to repurposing data for whole-cell modeling, foundational research is also needed to expand our experimental capabilities. While our capabilities to characterize the transcriptome and proteome have advanced rapidly over the past 20 years, our capabilities to characterize the metabolome, single cell variation, and temporal dynamics continue to lag. For example, additional capabilities to characterize the composition and dynamics of the metabolome could enable more complete flux balance analysis models.

\section*{The challenges to reusing data for whole-cell modeling}
While substantial data is already available for whole-cell modeling, unfortunately, most of this data is not readily accessible. The challenges to utilizing the existing data are several-fold. First, the existing data is distributed over a wide range of organisms and experimental conditions. As a result, only a small amount of data is available for each organism and experimental condition. One potential solution to this data sparsity is to leverage data from closely related organisms and conditions. However, few databases have been designed to help investigators search for such related data. Literature search engines such as Google Scholar and PubMed have also not been designed to help investigators find such related data. 

Second, our existing data is organized heterogeneously. Our existing data is scattered across many databases, as well as many individual journal articles. Additionally, the existing databases provide different interfaces and APIs. Furthermore, the existing data is described with many different formats, identifier systems, and ontologies. The effort required to deal with this heterogeneity distracts investigators from modeling. 

Third, many databases and articles only provide minimal metadata or minimally structured metadata. The lack of detailed metadata is part of why it is difficult to find measurements of related organisms and conditions. The lack of detailed, consistently structured metadata also makes it challenging to interpret and integrate data accurately.

Fourth, a significant amount of data is not available in any reusable form. Despite increasing emphasis on data sharing and reuse \cite{pmid26978244}, many results are still reported without their underlying data. One contributing factor is the lack of domain-specific formats and databases for many types of data. Such shared infrastructure makes it easier for authors to share data and easier for other investigators to reuse it. In the absence of such infrastructure, authors often have little incentive to share data, and reviewers often have low expectations for data sharing. Furthermore, with notable exceptions for genetic and structural data, many journals still have porous guidelines that permit publication without sharing the underlying data.

\section*{Emerging tools for sharing, discovering, and reusing data}
Efforts to make data easier to share, discover, and reuse for whole-cell modeling and other research are underway. This includes the development of standard formats and ontologies for describing data, central databases for storing data, and tools for discovering specific data. Here, we highlight some of the most relevant emerging resources for whole-cell modeling.

\subsection*{Formats for exchanging data for whole-cell modeling}
Three notable formats for capturing some of the data and knowledge needed for whole-cell modeling include the Investigation/Study/Assay tabular (ISA-Tab) format \cite{pmid20679334}, the Multicellular Data Standard (MultiCellDS) \cite{friedman2016multicellds}, and BioPAX \cite{demir2010biopax}. ISA-Tab is ideal for high-dimensional data, such as transcriptome-wide measurements of RNA turnover rates, which lack more specific formats. MultiCellDS is an emerging format intended to capture a digital ``snapshot'' of a cell line, encompassing measurements of its metabolome, transcriptome, proteome, and phenotype, as well as metadata about the environmental context of each measurement and the methods used to collect it. BioPAX is a format for describing knowledge about the molecules and molecular interactions inside cells. 

In our experience, whole-cell modeling requires both quantitative and relational data about multiple aspects of a cell. To capture this information for our first models, we developed the WholeCellKB schema \cite{karr2012wholecellkb}. Simultaneously, Lubitz and colleagues developed SBTab \cite{lubitz2016sbtab}, a tabular format with similar goals. As we began to explore additional models, we realized that many modelers both want to be able to use spreadsheets to quickly assemble datasets and use computer programs to quality control their datasets and incorporate them into models. To meet this need, we recently merged the concepts behind WholeCellKB and SBTab into ObjTables \cite{karr2020structured}, a set of tools that make it easy for modelers to use user-friendly spreadsheets to integrate data, define schemas for rigorously validating their data, and parse linked spreadsheets into data structures that are conducive to modeling. SEEK provides an online environment for managing datasets organized as spreadsheets \cite{wolstencroft2015seek}.

\subsection*{Formats for critical metadata for whole-cell modeling}
As we discussed above, structured metadata is critical for understanding and merging data. Because cells contain millions of distinct molecular species \cite{pmid29443976} due to combinatorial biochemical processes such as post-transcriptional and post-translational modification and complexation, we think it is particularly important for datasets to concretely describe the macromolecules that they characterize. Potential formats for describing polymers include BigSMILES \cite{pmid31572779}, BpForms/BcForms \cite{pmid32423472}, and HELM \cite{pmid22947017}. BpForms generalizes the IUPAC and IUBMB formats commonly used to describe unmodified DNAs, RNAs, and proteins to capture physiological polymers with modifications, crosslinks, and nicks. BcForms is a simple format for concretely describing complexes. BpForms and BcForms can both be used inside datasets, such as ObjTables spreadsheets. BigSMILES and HELM are more general formats for concretely describing large molecules with syntaxes that diverge from the IUPAC and IUBMB formats. Most other metadata needed for whole-cell modeling can be captured using identifiers and ontology terms. RightField \cite{wolstencroft2011rightfield} defines a convention for embedding ontology terms into spreadsheets.

\subsection*{Centralized knowledgebases of information for whole-cell modeling}
Because whole-cell modeling requires multiple types of data, we believe that centralized databases are also needed to help investigators find and obtain data. Three pioneering efforts to centralize data for modeling cells were the CyberCell Database (CCDB) for quantitative data about \textit{E. coli} \citep[*]{sundararaj2004cybercell}, EcoCyc for qualitative and relational information about \textit{E. coli} \citep[{**}]{keseler2017ecocyc}, and NeuronDB and CellPropDB for quantitative data about membrane channels, receptors, and neurotransmitters \citep[*]{crasto2007senselab}. EcoCyc continues to be a valuable resource, particularly for the development of genome-scale metabolic models \cite{latendresse2012construction}. GEMMER is a newer database that aims to facilitate models of \textit{S. cerevisiae} \cite{mondeel2018gemmer}.

More recent efforts to aggregate data for modeling have refined and expanded the concepts pioneered by the CCDB, CellPropDB, EcoCyc, NeuronDB, and others. One additional concept which we believe is essential is crowdsourcing. Crowdsourcing data aggregation addresses the problem that no single lab can curate the entire literature, and it can help avoid duplicate efforts by multiple researchers to curate similar data. Two exemplary resources that embody this philosophy are the Omics Discovery Index (OmicsDI) \citep[{**}]{perez2017discovering}, which provides a search engine to discover over 20 different types of quantitative molecular data curated by more than 20 different communities, and Pathway Commons \cite{cerami2010pathway}, which provides a search engine for information about molecular interactions curated by more than 22 groups of curators. To make it easy to contribute to OmicsDI and Pathway Commons, contributors only need to contribute a small amount of information about each dataset (OmicsDI) and pathway (Pathway Commons). However, this strategy pushes the onerous work of aggregating and normalizing data from the developers of these resources to their users.

To further help modelers obtain data for whole-cell modeling, we developed Datanator \citep[{**}]{roth2021datanator}, an integrated database of data for modeling the biochemical activity of a cell. Datanator builds on many of the ideas pioneered by the CCDB, OmicsDI, and other databases. Like OmicsDI, Datanator is a meta database that leverages the curation efforts and expertise of several primary databases. Like the CCDB, Datanator provides data in a consistent format that is convenient for modelers. In addition, Datanator provides a search engine tailored to the sparse nature of our existing data. This search engine can help modelers compensate for the absence of direct measurements with measurements of similar molecules, molecular interactions, organisms, or experimental conditions. The OpenWorm project \cite{sarma2018openworm} aims to develop a similar resource to accelerate their efforts to model \textit{Caenorhabditis elegans}.

\section*{Roadmap to data for whole-cell modeling}
\begin{figure}[t]
\centering
\includegraphics{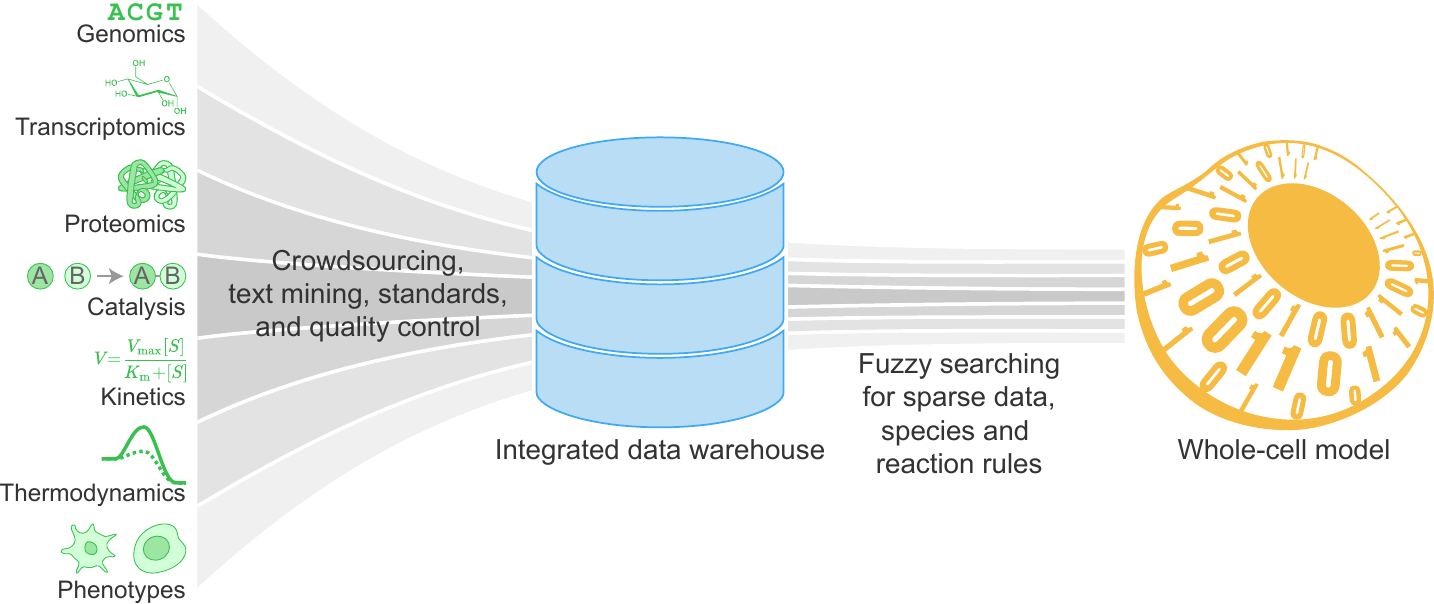}
\caption{\label{fig}\textbf{An integrated warehouse of molecular data and knowledge is needed to accelerate whole-cell modeling.} This warehouse could be assembled by combining multiple crowdsourced databases for different types of data with data automatically mined from the literature. Models could be systematically constructed from this warehouse using sets of rules that encode biochemical processes and physical laws.}
\end{figure}

Despite progress, we still only have a fraction of the data that will likely be needed for whole-cell modeling, and it remains tedious to gather the data that does exist. Ultimately, new experimental methods will be needed to fill the gaps in our understanding of the individual molecules and molecular interactions in cells. To enable investigators to independently train and test their models, increased automation will also be needed to generate data about a wider range of genotypes and environmental conditions. Most importantly, investigators need to pool their efforts so that everyone has access to more data. Here, we outline one way the community could work together to assemble the data that many modelers need \textcolor{blue}{(Figure~\ref{fig})}.

To facilitate the density of data needed for more comprehensive models, the community could first focus on a small number of organisms, cell types, and environmental conditions such as \textit{E. coli}, \textit{S. cerevisiae}, and \textit{H. sapiens} stem cells in rich media. Second, the community could coordinate the generation of different types of data to ensure that these cells are characterized deeply and avoid multiple researchers redundantly generating similar data. Third, the community could align on common formats and units for each type of data and for metadata, such as the genotype of each measured organism, the structure of each measured molecule, and the composition of each measured media condition. Fourth, databases such as Datanator, EcoCyc, OmicsDI, and Pathway Commons could be enhanced or merged into a single integrated system that can capture a wide range of qualitative, quantitative, and relational data. This infrastructure could be structured to both crowdsource the aggregation of data and enable separate curators to quality control the aggregated data. Text mining could also be used to automatically extract data from individual articles.

Once this data warehouse is available, additional methods and tools will be needed to use it to construct models. One possible way to use the data will be to devise rules, or templates, for generating species, reactions, rate laws, and rate parameters for specific types of data. For example, a rule could be created that generates protein species and translation and protein turnover reactions based on sequenced genomes, computed locations of start and stop codons, and measured protein abundances and half-lives. Such rules could encode biochemical processes such as translation and physical laws such as mass-action kinetics. Potentially, entire models could be constructed from such rules. This workflow would enable complex, detailed models to be systematically and transparently constructed from comparatively small sets of rules. We are building a system that will enable such rules. We anticipate it will accelerate the construction of large models.

\section*{Conclusions}
Despite the challenges to assembling the data needed for whole-cell modeling, we are confident that the combination of technology development, standardization, and collaboration outlined above will enable substantially more comprehensive, predictive, and credible models. Our Datanator database prototypes many of these ideas. To illustrate their potential, we are currently using Datanator to help construct a higher resolution model of the metabolism of \textit{E. coli}. To move forward, we encourage the community to join existing efforts to aggregate data such as Datanator, EcoCyc, and OmicsDI by helping to gather, integrate, or quality control data, or develop formats and tools that could facilitate these efforts.

\section*{Declaration of competing interest}
None.

\section*{Acknowledgments}
We thank Paul Lang, Zhouyang Lian, Wolfram Liebermeister, Saahith Pochiraju, Yosef Roth, and David Wishart for enlightening discussions about data for whole-cell modeling. This work was supported by the National Institutes of Health [grant numbers R35GM119771, P41EB023912].

\section*{References}
Papers of particular interest, published within the period of review, have been highlighted as:
\begin{itemize}[noitemsep, topsep=0pt, align=left, leftmargin=30pt, labelwidth=10pt, labelsep=4pt]
\item[*]  of special interest
\item[**] of outstanding interest
\end{itemize}


\end{document}